\documentclass[12pt]{article}
\usepackage{graphicx}
\usepackage{cite}
\usepackage{amssymb}
\usepackage{amsmath}
\usepackage[margin=1in]{geometry}
\usepackage{listings}
\usepackage{multirow}
\usepackage{booktabs} % for much better looking tables
\usepackage{array} % for better arrays (eg matrices) in maths
\usepackage{paralist} % very flexible & customisable lists (eg. enumerate/itemize, etc.)
\usepackage{verbatim} % adds environment for commenting out blocks of text & for better verbatim
\usepackage{subfig} % make it possible to include more than one captioned figure/table in a single float
\usepackage{textcomp} % provide euro and other symbols
\usepackage{soul}
\usepackage{algorithm}
\usepackage{algpseudocode}
\usepackage{amsfonts}
\usepackage{url}
\usepackage{float}
\usepackage{color}

\newcommand{\bfx}{\mathbf{x}}

\newcommand{\R}{{\mathbb R}}

\title{\bf Time-series imputation using low-rank matrix completion}
\author{Thomas Poudevigne%
\thanks{\texttt{tompoudevigne@gmail.com}}
\ and Owen Jones%
\thanks{\texttt{joneso18@cardiff.ac.uk}}}
\date{August 5, 2024}
\begin{document}
\maketitle
\begin{abstract}
We investigate the use of matrix completion methods for time-series imputation.
Specifically we consider low-rank completion of the block-Hankel matrix representation of a time-series.
Simulation experiments are used to compare the method with five recognised imputation techniques with varying levels of computational effort.
The Hankel Imputation (HI) method is seen to perform competitively at interpolating missing time-series data, and shows particular potential for reproducing sharp peaks in the data.
\end{abstract}

\section{Introduction}
Imputing missing data in time-series is challenging because of the temporal dependency between observations.
Here we develop an imputation method using low-rank matrix completion of a block-Hankel matrix representation of the time-series.
We compare its effectiveness to five recognised imputation techniques with varying levels of computational effort.

Let $\bfx(1), \ldots, \bfx(n)$ be a time-series in $\R^d$.
Fix a lag $k$, then taking the $\bfx(i)$ as row vectors, we restructure the data as a block-Hankel matrix\cite{golyandina2001analysis} of size $(n-k+1) \times (kd)$:
\[
H = \left(
\begin{array}{cccc}
\bfx(1) & \bfx(2) & \cdots & \bfx(k) \\
\bfx(2) & \bfx(3) & \cdots & \bfx(k+1) \\
\vdots & \vdots & \ddots & \vdots \\
\bfx(n-k+1) & \bfx(n-k+2) & \cdots & \bfx(n)
\end{array}
\right)
\]

Using $H$ we can recast time-series imputation as a matrix completion problem.
From the block-Hankel construction, and the assumed temporal dependence of the $\bfx(i)$, it is natural to suppose that $H$ is low-rank.

Let $M$ be an $m \times n$ matrix and $\Omega \subset \{1,\ldots,m\} \times \{1,\ldots,n\}$ the indices of its non-missing values, then the low-rank completion of $M$ solves
\begin{equation}\label{eqn1}
\begin{array}{c}
\min_X \mbox{ rank}(X) \\[6pt]
\mbox{s.t. } X(i,j) = M(i,j) \mbox{ for all } (i,j) \in \Omega.
\end{array}
\end{equation}

Solving (\ref{eqn1}) is not easy, so in practice more tractable relaxations are used.
The nuclear norm of $X$ is the sum of its singular values, written $\|X\|_*$.
%\footnote{The square roots of the eigenvalues of $X'X$.}
On the set of matrices with spectral norm at most one, 
%\footnote{The absolute value of the largest eigenvalue.}
the nuclear norm is the convex envelope of the rank operator, so the nuclear norm gives a natural relaxation of the problem.
Cand\`es \& Recht \cite{candes2008exact} showed that for a large class of matrices with entries missing independently and uniformly---given certain conditions on the rank of $M$ and the number of missing values---with high probability the problem
\begin{equation}\label{eqn2}
\begin{array}{c}
\min_X \|X\|_* \\[6pt]
\mbox{s.t. } X(i,j) = M(i,j) \mbox{ for all } (i,j) \in \Omega
\end{array}
\end{equation}
has a unique solution equal to $M$.

For a block-Hankel matrix the missing values are certainly not missing independently, so Cand\`es \& Recht's result doesn't apply.
Dai \& Pelckmans \cite{dai2015nuclear} and Usevich \& Comon \cite{usevich2016hankel} have some results for Henkel matrices when the missing values are in the bottom right corner (as for a one-dimensional forecasting problem), but these don't help either.
We will use the nuclear norm relaxation none-the-less, though of course our results are all experimental.

The nuclear norm is convex (since it's a norm), and the constraint is linear (on the space of $m \times n$ matrices), so our matrix completion problem (\ref{eqn2}) is a convex optimisation problem, and can in principal be solved using interior point methods.
In practice we reformulate (\ref{eqn2}) as a semi-definite programme, see for example \cite{recht2010guaranteed}:
\begin{equation}\label{eqn3}
\begin{array}{c}
\min_{W_1, W_2, X} \mbox{trace}(W_1) + \mbox{trace}(W_2) \\[6pt]
\mbox{s.t. } X(i,j) = M(i,j) \mbox{ for all } (i,j) \in \Omega \\[6pt]
\left[\begin{array}{cc} 
W_1 & X \\
X^T & W_2
\end{array} \right]
\succcurlyeq 0.
\end{array}
\end{equation}
Here $W_1$ and $W_2$ are symmetric $m\times m$ and $n\times n$ matrices.
By vectorising the matrices $W_1$, $W_2$ and $X$ this can be expressed as a linear programme on a cone, and can be solved using, for example, the splitting conic solver of O'Donoghue et al.\cite{ocpb:16,scs}, which can accommodate high dimensional problems.
Alternatively Cai, Cand\`es \& Shen \cite{cai2010singular} have a singular value thesholding method that also works for large matrices.
Both of these solution techniques are iterative, and the equality constraint $X(i,j) = M(i,j) \mbox{ for all } (i,j) \in \Omega$ is effectively replaced by the inequality constraint
\[
\sum_{(i,j) \in \Omega} (X(i,j) - M(i,j))^2 \leq \epsilon^2,
\]
for some small tolerance $\epsilon$.

Previously Butcher \& Gillard\cite{butcher2017simple} gave an example using low-rank matrix completion for imputation, but without any comparative analysis.
In related work Gillard \& Usevich\cite{gillard2018structured} gave some theoretical results on the use of low-rank matrix completion to forecast time-series, together with some numerical studies.

\section{Experimental setup}\label{hidatas}
\begin{figure}
\centering
\includegraphics[width=\textwidth]{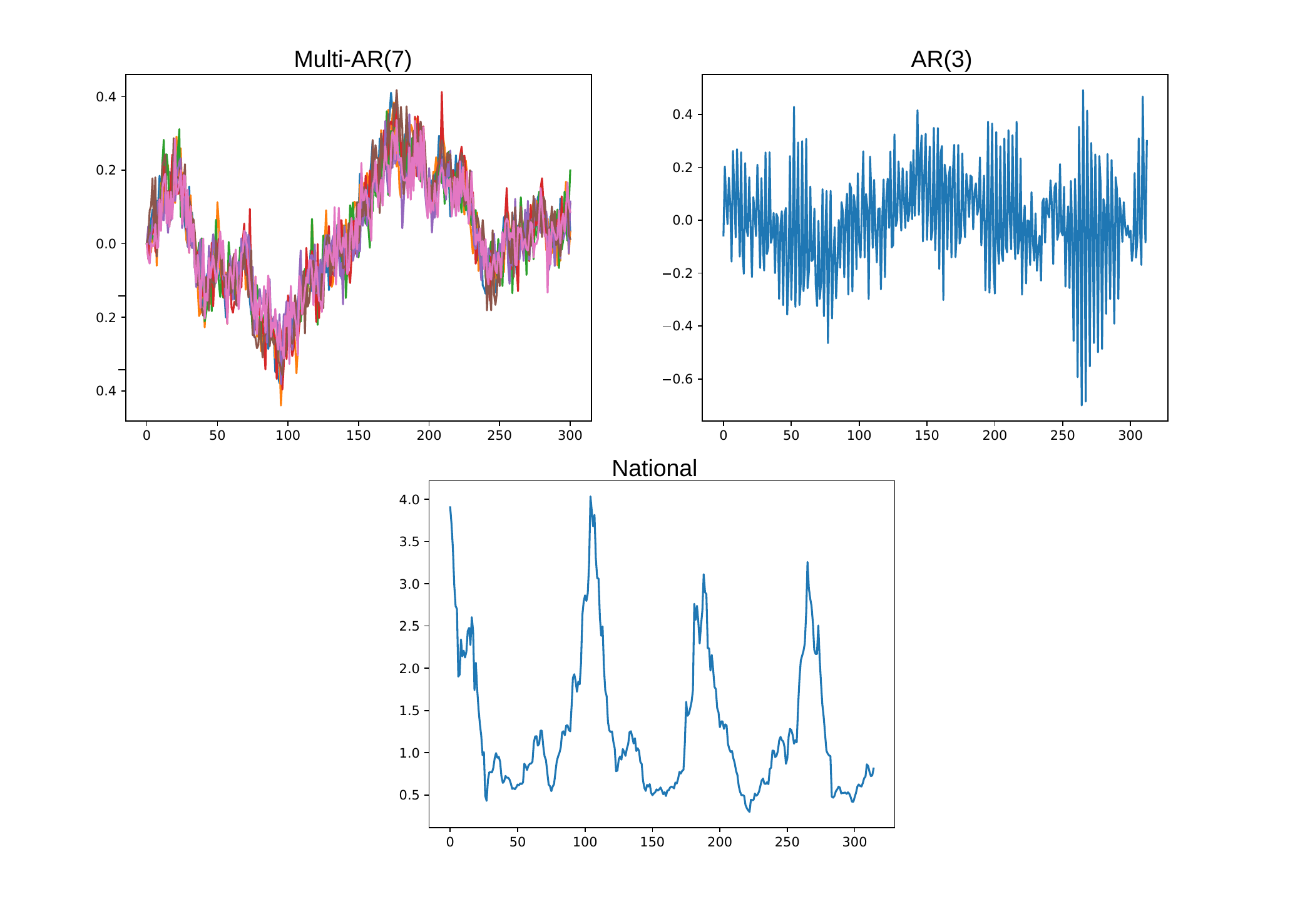}
\caption{The sample data used for experiments.
Clockwise from the top-left they are from the VAR(1) model, the AR(3) model, and the Welsh Government Wastewater Surveillance programme SARS-CoV-2 levels (WC). The different variables of the VAR(1) sample are given by different colours.}
\label{setsofar}
\end{figure}

We will refer to our method as \textit{Hankel Imputation} (HI) in what follows.
We tested the HI method with two artificial and one real data set:
\begin{description}
\item[VAR(1)] A lag 1 dimension 7 vector auto-regression model with parameters
\[
\begin{pmatrix}
0.6&0.22&0.13&0.02&0.05&0.003&0.0004\\
0.6&0.12&0.19&0.03&0.03&0.004&0.00041\\
0.5&0.15&0.12&0.07&0.04&0.007&0.00042\\
0.6&0.13&0.19&0.04&0.03&0.003&0.00043\\
0.4&0.122&0.15&0.07&0.02&0.001&0.00044\\
0.55&0.162&0.17&0.13&0.03&0.0045&0.00045\\
0.45&0.152&0.12&0.07&0.01&0.0082&0.00046\\
\end{pmatrix}.
\]
A single sample of length 300 was used, see Figure \ref{setsofar}.
\item[AR(3)] A lag 3 auto-regression model with parameters $(0.1,-0.3,0.9)$.
A single sample of length 300 was used, see Figure \ref{setsofar}.
\item[WC] SARS-CoV-2 levels in wastewater collected by the Welsh Government Wastewater Surveillance programme.
A time-series of 315 daily data points was extracted from the national signal, see Figure \ref{setsofar}. 
Each point gives the level of SARS-CoV-2 gene copies per 100,000 people
across 47 wastewater treatment works covering 75\% of the Welsh population\cite{w14182885}.
\end{description}
All three time-series are complete.
We generated series with 10\%, 20\%, ..., 70\% missing values by removing values uniformly at random.
This was always done in a nested fashion, so that the values missing at the 10\% level are also missing at the 20\% level, and so on.

The HI method was compared with 5 commonly used univariate imputation methods.
\begin{description}
\item[Linear] Linear interpolation.
\item[Spline] Spline interpolation.
\item[Stine] Stineman interpolation.
\item[Kalman] Interpolation using a structural model and Kalman smoothing.
\item[EWMA] Interpolation using an Exponential Weighted Moving Average.
\end{description}
We used the implementations from the R package \texttt{imputeTS}\cite{moritz2017imputets}, with the default parameter settings.

\subsection{Quantifying performance}\label{tan}
We are interested in how the HI method picks up high-frequency (noise) signal components as well as low-frequency (trend) components.
Accordingly, rather than measuring the performance using the mean squared error, we used used two performance measures designed to measure separately how well the method imputes the trend and noise components of the time-series.

It is common to decompose a time-series into seasonal, trend and noise components, see for example \cite{PETROPOULOS2022705}.
Our VAR(1) and AR(3) series are stationary and have no seasonal component, and the WC series is only observed over a single year so its seasonal component can't be distinguished from trend, so we considered only trend and noise components.

Let $x(1), \ldots, x(n)$ be a univariate time-series (with no seasonal component), then we estimated the trend $x_\tau(n)$ using a simple smooth with a moving window of length $2m+1$, for some user chosen $m$:
\begin{equation}\label{smooth.eqn}
x_\tau(t) = \frac{1}{2m+1} \sum_{s=k-m}^{k+m}x(s).
\end{equation}
The noise term is then just the remainder:
\[
x_\epsilon(t) = x(t) - x_\tau(t).
\]
Suppose that observations $t_1, \ldots, t_m$ are considered missing, and let $y(t_1), \ldots, y(t_m)$ be the corresponding imputed values.
For $s \not\in \{t_1, \ldots, t_m\}$ put $y(s) = x(s)$, then our measure for how well the imputed data fits the trend is
\[
T = \sqrt{\textstyle \frac{1}{m} \sum_{i=1}^{m}(x_\tau(t_i)-y_\tau(t_i))^2}
\]
and our measure for how well it fits the noise is
\[
E = \left|
\sqrt{\textstyle \frac{1}{m} \sum_{i=1}^{m} x_\epsilon(t_i)^2} - 
\sqrt{\textstyle \frac{1}{m} \sum_{i=1}^{m} y_\epsilon(t_i)^2} \right|.
\]
We refer to these as the Trend Score and Noise Score. 
In both cases the smaller the better.
For the multivariate VAR(1) dataset the scores were averaged over all variables.

\subsection{Parameter tuning}\label{tuning}
The HI method has two tuning parameters, the lag $k$ used to form the block-Hankel matrix, and the tolerance $\epsilon$ used when solving (\ref{eqn3}).

\begin{figure}
\centering
\includegraphics[width=0.6\textwidth]{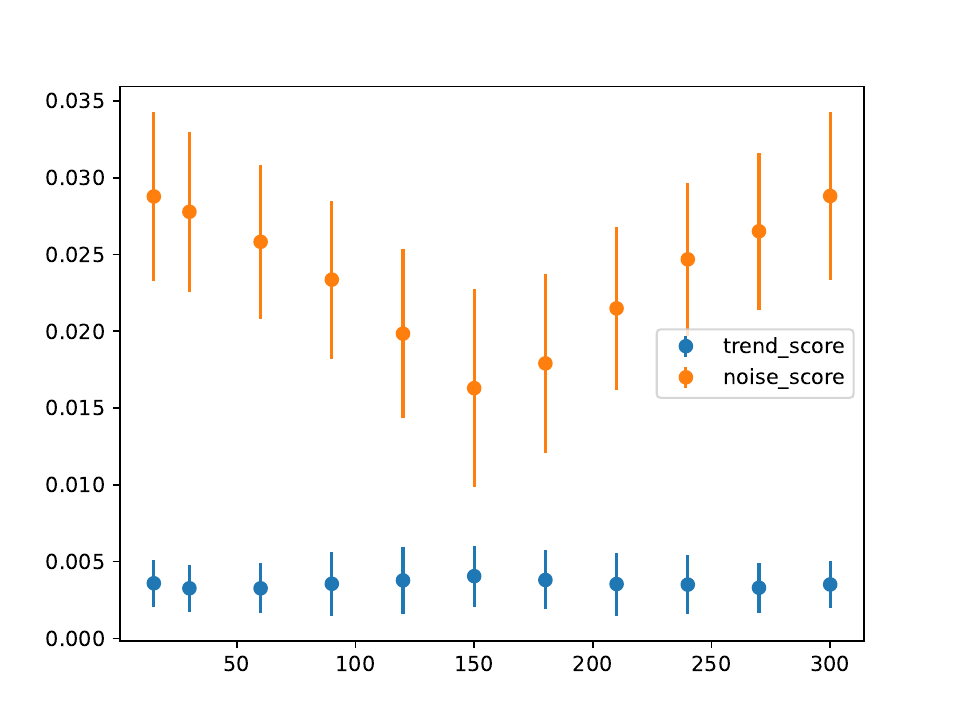}
\caption{The effect on performance of the lag $k$ used to form the block-Hankel matrix.
Results are for the WC data with 40\% missing data.
$\epsilon$ was fixed at $0.1$.
Each point gives the mean and 95\% confidence interval from 10 trials using independent choices of missing points.}
\label{lagmat}
\end{figure}

To test the choice of $k$ we used the WC data with 40\% missing data.
The Trend and Noise Score for different choices of $k$ are given in Figure \ref{lagmat}.

We see that the lag has some impact on the Noise Score but very little impact on the Trend Score. 
The time-series used has length 300 and dimension 1, so the lag $k = 150$ is the case where the block-Hankel matrix is approximately square, and this is when the HI method performs the best.
This agrees with Gillard \& Usevich\cite{gillard2018structured}, who observed that their forecasting method performed best when using square(ish) Hankel matrices. 
Accordingly in what follows we will always use lag 
\begin{equation}\label{specialk}
k = \lceil {(n+1)}/{(d+1)} \rceil.
\end{equation}

\begin{figure}
\centering
\includegraphics[width=0.6\textwidth]{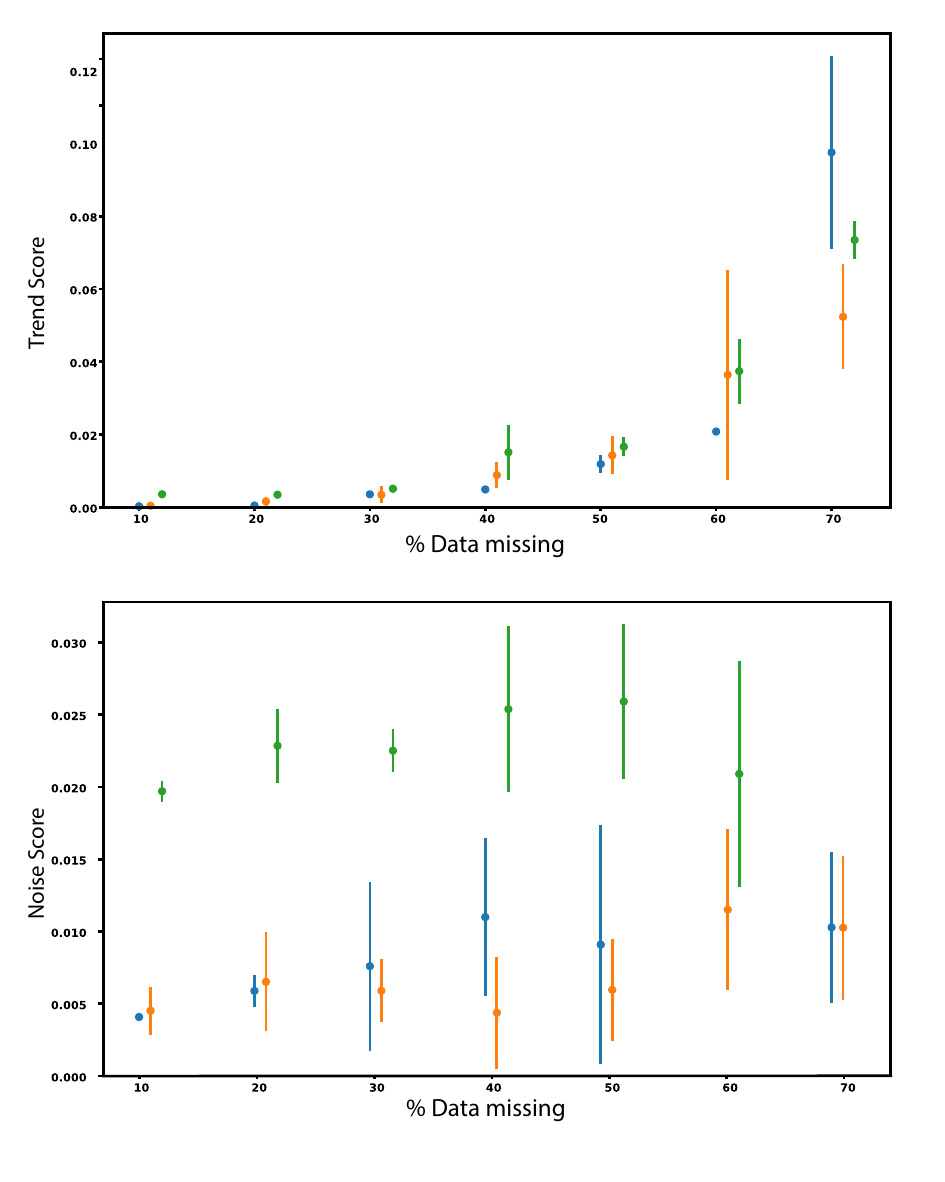}
\caption{The effect on performance of the tolerance parameter $\epsilon$.
Results for $\epsilon=1$ are in green, $\epsilon=0.1$ in orange, and $\epsilon=0.01$ in blue.
We used the Welsh Covid (WC) data set, and for each point we have the mean and 95\% confidence interval from 100 runs with independent sets of missing points.}
\label{expsparam}
\end{figure}

To test the effect of the tolerance $\epsilon$ we used the WC data set, with different levels of missing data.
The results are given in Figure \ref{expsparam} and we see that in general for the best performance we want to take $\epsilon$ as small as possible.
In particular the Noise Score is sensitive to the choice of $\epsilon$: taking it too large can result in a significant drop in performance.
However smaller $\epsilon$ comes at the cost of increased run-time to solve (\ref{eqn3}), so there is a trade-off between accuracy and resource costs.

For our experiments we took $\epsilon = 0.01$ to give a balance between accuracy and resource costs. 

\subsection{Blocking}
Suppose our original time-series has length $n$.
Choosing $k$ according to (\ref{specialk}) gives us a block-Hankel matrix $H$ of size $O(n) \times O(n)$.
To calculate the nuclear norm we need the singular value decomposition of $H$, which has a computation cost $O(n^3)$.
This can be prohibitive for large $n$, in which case we can reduce the computational burden by splitting the time-series into blocks and applying the HI method separately to each block.

Consider splitting the time-series into two blocks of length $n/2$.
Compared to the original series, the computational cost for each block would reduce by a factor $2^{-3}$, so the overall cost would reduce by $2^{-2}$, that is 25\% of the original cost.
Of course solving (\ref{eqn3}) requires more than a singular value decomposition, so this should only be taken as a guide.

\begin{figure}
\centering
\includegraphics[width=0.6\textwidth]{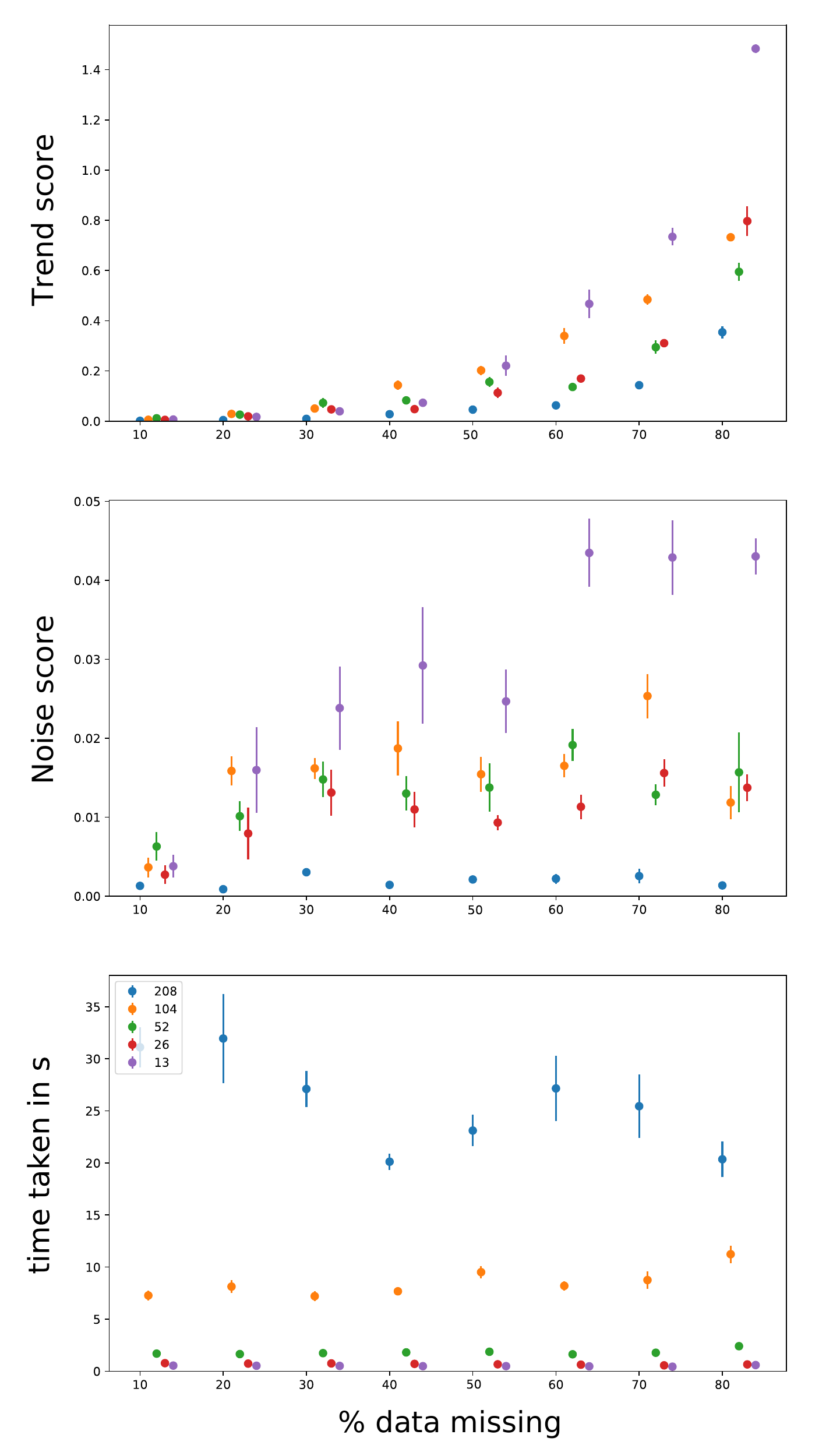}
\caption{The effect of block size on performance and computation time, using the Welsh Covid (WC) data set.
The top panel gives the Trend Score squared, the middle panel the Noise Score, and the bottom panel the computation time.
Each point is the average of 10 experiments using different sets of missing values and the vertical bars give 95\% confidence intervals.}
\label{allg1}
\end{figure}

A simulation experiment was used to quantify the effect of block calculations on computational time and performance.
We used 208 observations from the Welsh Covid (WC) data, with missing observations chosen uniformly at random at the 10\%, 20\%, ..., 80\% level.
For each of these data sets imputation was carried out using batches of size 208, 104, 52, 26 and 13. 
The results are given in Figure \ref{allg1}.
Computation time depends mainly on the block size, and scales roughly as suggested by our calculation above.
At all levels of missingness there is a loss in performance when using smaller block sizes---in particular for the Noise Score---which becomes more marked as the level of missingness increases.

For our experimental results below we used a single block.

\section{Results}\label{hires}

\begin{figure}
\centering
\includegraphics[width=0.495\textwidth]{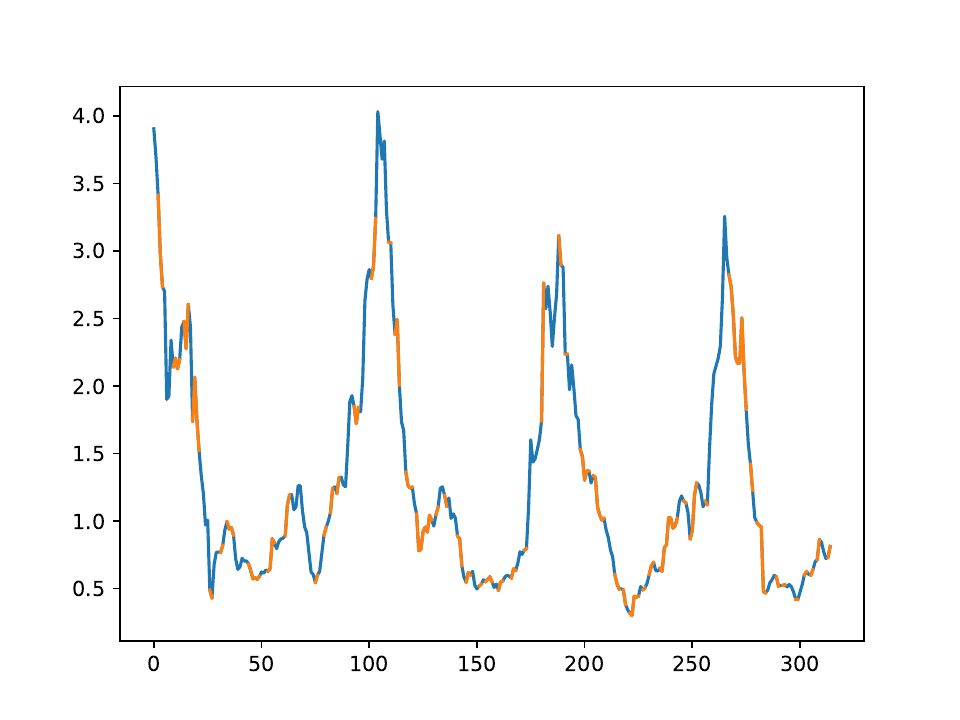}
\includegraphics[width=0.495\textwidth]{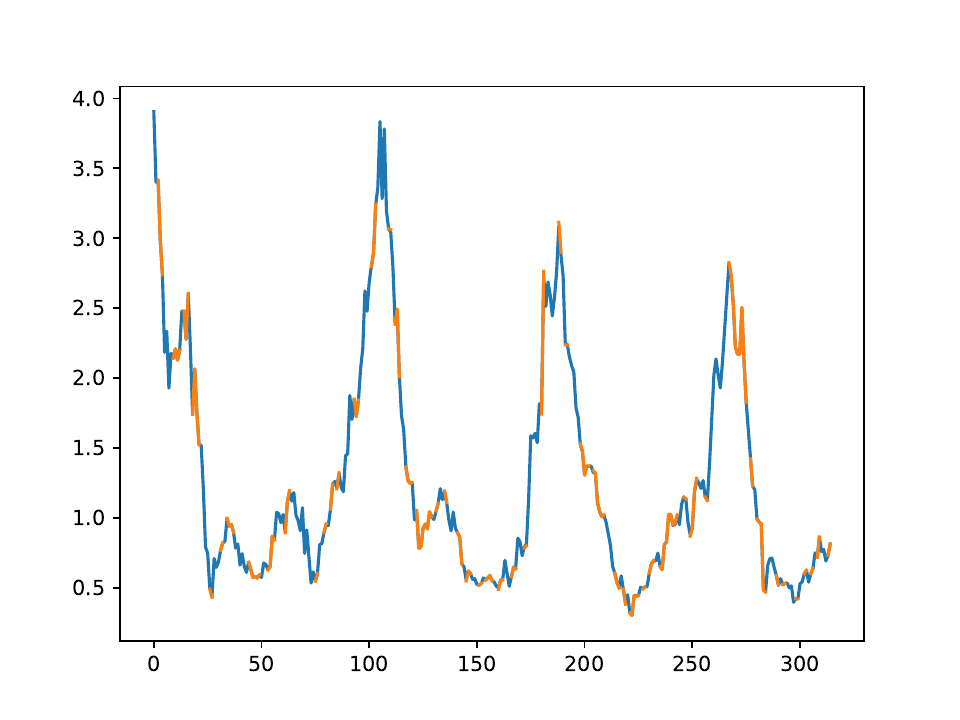}
\caption{Imputing missing values from the Welsh Covid (WC) data set using the HI method.
On the left we have the original time series with the values removed shown in blue 
On the right the blue values have been replaced by values imputed using the HI method.}
\label{dhisamle}
\end{figure}

We first illustrate the HI method using the Welsh Covid (WC) data.
40\% was removed uniformly at random, then the HI method was applied using $k$ according to (\ref{specialk}) and $\epsilon = 0.01$.
The results are given in Figure \ref{dhisamle} and we see that qualitatively the method is doing a good job of picking up both trend and noise when imputing the missing values.

\begin{figure}
\centering
\includegraphics[width=\textwidth]{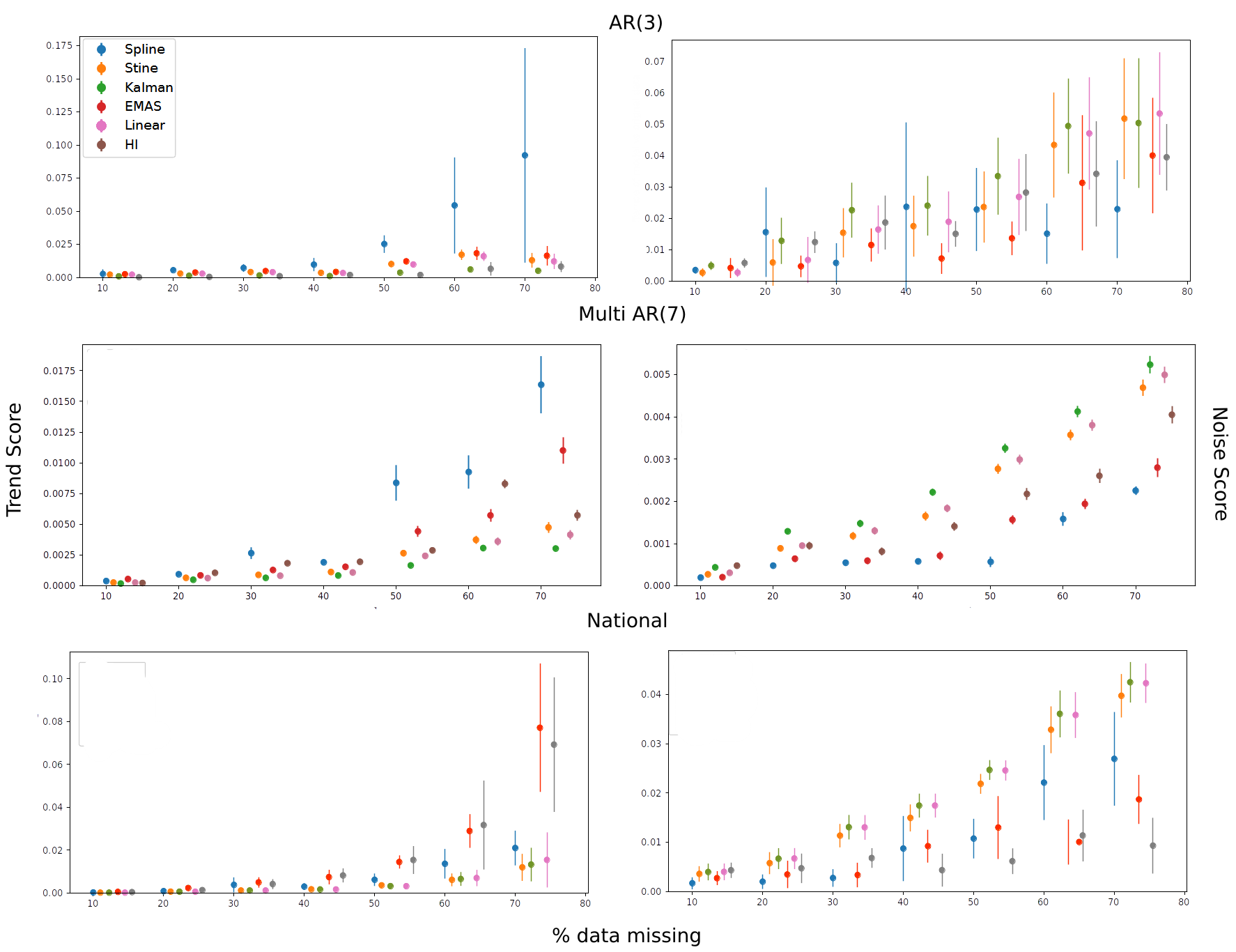}
\caption{Performance of the HI method (in grey) and five comparators: linear interpolation in pink, spline interpolation in blue, Stine interpolation in orange, Kalman smoothing in green, and exponential weighted moving average in red.
Results for the AR(3), VAR(1) and WC data sets are given in rows 1 to 3 respectively.
For each level of missingness the Noise Score and Trend Score were calculated using {10} independent sets of missing values, and we have plotted the mean and 95\% confidence interval of these.}
\label{allg}
\end{figure}

Our quantitative results are given in Figure \ref{allg}.
To calculate the trend used in the Trend Score (\ref{smooth.eqn}) we used smooths of radius 7, 3 and 7 respectfully for the VAR(1), AR(3) and WC data sets.
For the AR(3) dataset the Trend Score of the HI method was the lowest, and the Noise Score was low though not the lowest. 
For the VAR(1) dataset the Trend Score of the HI method was low but the Noise Score only average compared to the other methods. 
Finally, for WC dataset the Noise Score of the HI method was the lowest but the Trend score was high. 

Overall the performance of the HI method is visually very good and quantitatively it performed well compared to the other methods considered.
On average the HI method was better at getting the noise right than the other methods we looked at.
The Welsh Covid (WC) data is qualitatively different to the AR(3) and VAR(1) data in that it cycles through successive sharp peaks.
Traditional interpolation methods find it very hard to impute missing peaks, however visually the HI method seems to do very well in this regard, and this is backed up by the low Noise Scores for this dataset.
In conclusion the HI method seems to perform competitively at interpolating missing time-series data, and shows particular potential for reproducing sharp peaks in the data.

\paragraph{Acknowledgements}
T.P.\ was funded by the KESS2 East programme of the European Commission and by D\^wr Cymru Welsh Water. Data used were supplied by the Welsh Government under the Welsh Wastewater Programme (C035/2021/2022) in partnership with Cardiff University and Bangor University.

\bibliographystyle{abbrv}
\bibliography{brg.bib}
\end{document}